\documentclass[conference]{IEEEtran}
\IEEEoverridecommandlockouts
\usepackage{cite}
\usepackage{amsmath,amssymb,amsfonts}
\usepackage{algorithmic}
\usepackage{graphicx}
\usepackage{textcomp}
\usepackage{xcolor}
\usepackage{multirow}
\def\BibTeX{{\rm B\kern-.05em{\sc i\kern-.025em b}\kern-.08em
    T\kern-.1667em\lower.7ex\hbox{E}\kern-.125emX}}
\begin{document}

\title{Authorship Identification of Source Code Segments
Written by Multiple Authors
Using Stacking Ensemble Method\\
}

\author{\IEEEauthorblockN{Parvez Mahbub}
\IEEEauthorblockA{\textit{CSE Discipline} \\
\textit{Khulna University}\\
Khulna, Bangladesh \\
this@parvezmrobin.com}
\and
\IEEEauthorblockN{Naz Zarreen Oishie}
\IEEEauthorblockA{\textit{CSE Discipline} \\
\textit{Khulna University}\\
Khulna, Bangladesh \\
nazzarreen05@gmail.com}
\and
\IEEEauthorblockN{S.M.\ Rafizul Haque}
\IEEEauthorblockA{\textit{CSE Discipline} \\
\textit{Khulna University}\\
Khulna, Bangladesh \\
rafizul@cse.ku.ac.bd}
}

\maketitle

\begin{abstract}
    Source code segment authorship identification is the task of identifying the author of a source code segment
    through supervised learning.
    It has vast importance in plagiarism detection, digital forensics, and several other law enforcement issues.
    However, when a source code segment is written by multiple authors, typical author identification methods no
    longer work.
    Here, an author identification technique, capable of predicting the authorship of source code segments,
    even in case of multiple authors, has been proposed which uses stacking ensemble classifier.
    This proposed technique is built upon several deep neural networks, random forests and support vector
    machine classifiers.
    It has been shown that for identifying the author-group, a single classification technique is no longer
    sufficient and using a deep neural network based stacking ensemble method can enhance the accuracy significantly.
    Performance of the proposed technique has been compared with some existing methods which only deal with the
    source code segments written exactly by a single author.
    Despite the harder task of authorship identification for source code segments written by multiple authors,
    our proposed technique has achieved promising results evident by the identification accuracy,
   compared to the related works which only deal with code segments written by a single author.
\end{abstract}


\section{Introduction}\label{sec:introduction}
Source code segment author identification is a major research topic in the field of software forensics.
It has many uses such as plagiarism detection, law enforcement, copyright infringement
etc.\cite{Frantzeskou06,Kothari}.
Frantzeskou\cite{Frantzeskou05} mentioned that source code author identification is useful against cyber attacks in
the form of viruses, trojan horses, logic bombs, fraud, credit card cloning, and authorship disputes or proof of
authorship in court.
There are certain patterns that developers sub-consciously reflect in their codes based on their particular coding
style while still following the guidelines, standards, rules, and grammars of a language or framework\cite{Kothari}.
These pieces of information can be used to identify the author of the source code segment.

In recent years, open source software development has entered a new era.
A lot of big companies like Google, Microsoft, and many others are maintaining their projects open source.
Alongside, small and mid-level projects are being written by a group of authors.
In these cases, trivial author identification schemes no longer work.
When someone contributes to an open source project, the writing style of the original author of the source code
segment is no longer unique and it makes the author identification task harder.
Even worse case is when a project is equally contributed by a number of authors.
The writing style is then the aggregation of all the authors.
We aimed to solve this problem and proposed an approach to identify the author of a source
code even when it is contributed by more than one author.
In this paper, we have proposed an author identification technique using a stacking ensemble method composed of several deep
neural networks(DNN), random forests and support vector machines(SVM).

\subsection{Problem Definition}\label{subsec:problem-definition}
Authorship identification is the task of having some samples of code for several programmers and determining the
likelihood of a new piece of code having been written by each programmer\cite{Gray}.
As the name suggests, authorship identification of source code segments written by multiple authors is identifying
the author-label when the number of authors of source code segment is more than one.
This can be of two types.
One is the source code segment can be written by mostly one author and have small contributions from several other
authors.
Another is a source code segment can be directly written by a group of authors and have a roughly equal contribution
from each of them.
Both of these happen in open-source software and projects which are very popular nowadays.

\subsection{Motivation}\label{subsec:motivation}
Authorship identification of source code segment has a vast application area including plagiarism detection,
authorship dispute, software forensics, malicious code tracking, criminal prosecution, software intellectual
property infringement, corporate litigation, and software maintenance\cite{Yang,Tennyson,Frantzeskou06,Lange,Zhang,Mirza}.
In the case of authorship dispute, authorship identification can be a solution.
Given the source code segment and the candidate owners, the likelihood of each candidate of being the author of the
source code can be determined\cite{Kothari}.
Again, Kothari\cite{Kothari} identified that author identification is useful for detecting the author of the
malicious code.
Software companies can also use authorship identification system to keep track of programs and modules for better
maintenance\cite{Zhang}.
Though source code segments are much more restrictive and formal than spoken or written language, they inhibit a
large degree of flexibility\cite{Frantzeskou05}.
According to Shevertalov\cite{Shevertalov}, using differences in the way programmers express their idea, their
programming style can be captured.
This programming style, in turn, can be used for author identification.
Although, a large number of works already done regarding source code segment author identification, according to
Frantzeskou\cite{Frantzeskou06}, the future of source code segment author identification is in collaborative
projects to which we aimed at.

The remaining sections are organized as follows.
Section~\ref{sec:background-study} contains a briefing on background topics regarding this work.
Section~\ref{sec:related-works} contains a summary of the related works.
In section~\ref{sec:methodology}, we discuss our author identification technique for multiple authors.
In section~\ref{sec:experimental-results}, the experimental results of our proposed technique are analyzed
and compared with that of some related works.
Finally, in section~\ref{sec:conclusion}, the conclusion is stated with possible future  direction of this work.

\section{Background}
\label{sec:background-study}
\subsection{Ensemble Method}\label{subsec:ensembling-method}
By combining several methods, ensembling method helps to improve the results of machine learning.
An ensemble is often more accurate than any of the single classifiers in the ensemble.
According to Maclin\cite{Maclin}, an ensemble consists of a set of individually trained classifiers whose
predictions are combined, while classifying instances, by the ensemble method.
These meta-algorithm combines several machine learning techniques into one predictive model.
In our work, we used stacking ensemble in order to improve our prediction performance.

\subsection{Random Forests}\label{subsec:random-forests}
Random forest is an ensemble learning method where each classifier in the ensemble is a decision tree classifier.
This collection of classifiers is called a forest.
During classification, each of the decision trees gives its vote and the result is based on the majority of the
votes.

\section{Related Works}
\label{sec:related-works}
Numerous works are available on source code segment author identification using a variety of features and classifiers.
However, very few of them use machine learning techniques to identify the author of source code segments.

According to \v Dura\v c\'{\i}k\cite{Duracik}, there are several approaches to identify the author of source code segment.
The first one is text-based and uses plain text as an input.
The second level is token or metric based.

\subsection{Text Based Approaches}
\label{subsec:text-based-approach}

The first approach, which treats source code segment as plain text, is a form of natural language processing.
This approach cannot make use of the programmatic structure of source code segment.

Frantzeskou et al.\cite{Frantzeskou06} proposed a technique called Source Code Author Profiles(SCAP) for author
identification.
They generated byte level n-gram author profile and compared with previously calculated author profiles.
Burrows\cite{Burrows12} mentioned, the SCAP method truncates the author profiles that are greater than the maximum
profile length causing a bias towards the truncated profiles.

Burrows et al.\cite{Burrows} proposed an approach using information retrieval.
They generated n-gram tokens from the source code segments and indexed them in a search engine to query the author of source
code and return a ranking list of authors which matched the n-gram token of the source code segment with 67\% accuracy.

\subsection{Metric Based Approaches}
\label{subsec:metric-based-approaches}

Frantzeskou\cite{Frantzeskou06} pointed out that metric-based author identification is divided into two steps.
The first step is extracting the code metrics that represent the author's style and the second part is using those
metrics to generate a model that is capable of labeling a source code segment by corresponding author name.
However, a large amount of time is required to gather all possible metrics and examine to choose only the metrics
responsible for differing the authors' style.

Lange and Spiros\cite{Lange} assumed that the code metrics histogram should vary from author to author as of their
coding style.
From a number of source code metrics, an optimum set was selected using genetic algorithms(GA) and used as input for
the nearest neighbor(NN) classifier.
This method achieved 55\% accuracy.
According to Yang\cite{Yang}, some of the features of this paper are unbounded.
For example, the indentation category.

Shevertalov el al.\cite{Shevertalov} proposed a technique based on GA\@.
The metrics are extracted from the source code segment to make a histogram which is sampled using GA\@.
The author profile is produced using categorized histogram samples.
For files, they achieved 54\% accuracy and for projects, they achieved 75\% accuracy.
Yang\cite{Yang} mentioned that the details of the final feature set are not mentioned in this paper.
So, the feature set is non-reproducible.

Bandara and Wijayarathna\cite{Bandara} used the deep Neural Network for source code segment author identification.
The converted source code metrics they used to feed a neural network are identical to that of Lange et al.\cite{Lange}.
Their deep neural network consisted of three restricted Boltzmann machine (RBM) layers and one output layer.
They achieved 93\% accuracy.

Zhang et al.\cite{Zhang} used SVM to identify the author of source code segment.
They categorized their feature into four groups namely -- programming layout feature, programming style feature,
programming structure feature and programming logic feature.
They used sequential minimal optimization(SMO) as the classifier for SVM\@ and achieved
98\% and 80\% accuracy for two different datasets.

\section{Author Identification of Source Codes Written by Multiple Authors}
\label{sec:methodology}

Our developed author identification approach consists of four phases.
Firstly, source code metrics are extracted from the source code segments in the training set.
These extracted metrics are then converted to feature vectors.

Secondly, these feature vectors are fed to five individual base classifiers along with corresponding class-labels
to train the author signatures to the base classifiers.
In the case of open source contribution, class-label means the owner of the source code segment and in the case of a group
of authors, the whole group is considered as the class label.
By author signature, the coding style of a particular class-label is meant.
Caruana\cite{Caruana08} showed that, in general, for classification problem, random forest, DNN, decision tree, and
SVM are the top four algorithms.
Hence, our chosen classifiers are DNN, random forest with CART decision trees\cite{CART}, random forest with C4.5
decision trees\cite{C4.5}, $C$-SVM and $\nu$-SVM\@.

\begin{figure}[ht]
    \centering
    \includegraphics[scale=.35, angle=0]{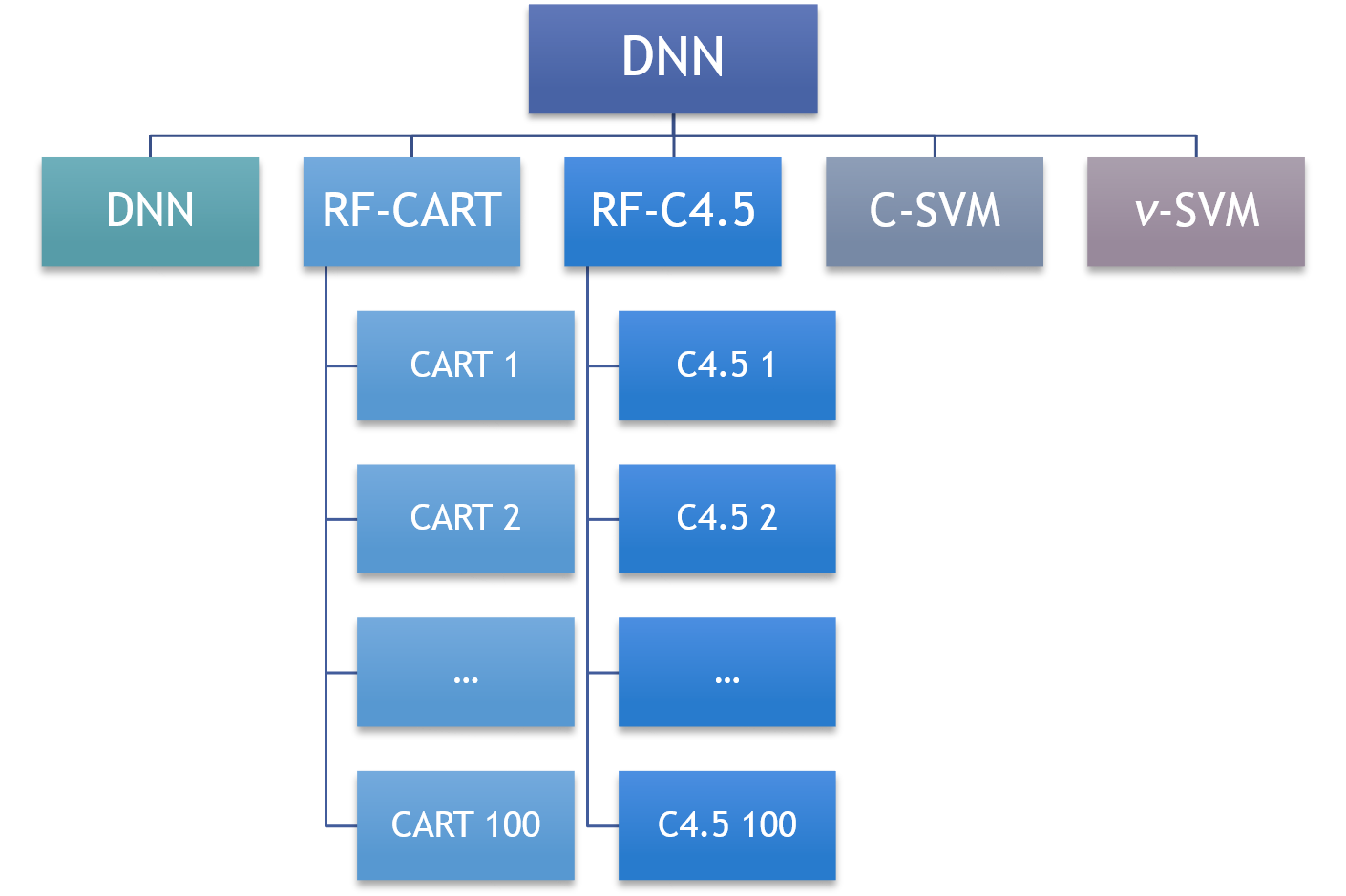}
    \caption{Block diagram of the architecture of the stacking ensemble method}
    \label{fig:clf-arch}
\end{figure}

Thirdly, each of the classifiers generates the posterior class-probability according to their predictions.
These outputs are called meta-features.
Meta-features are used as the input for a meta-classifier.
Then the meta-classifier is trained based on the meta-features and output.
This approach is known as stacking ensemble.
Another deep neural network is used as the meta-classifier.
Figure~\ref{fig:clf-arch} shows a block diagram of the architecture of the stacking ensemble method we have
designed.

\begin{figure}[ht]
    \centering
    \includegraphics[scale=.35, angle=0]{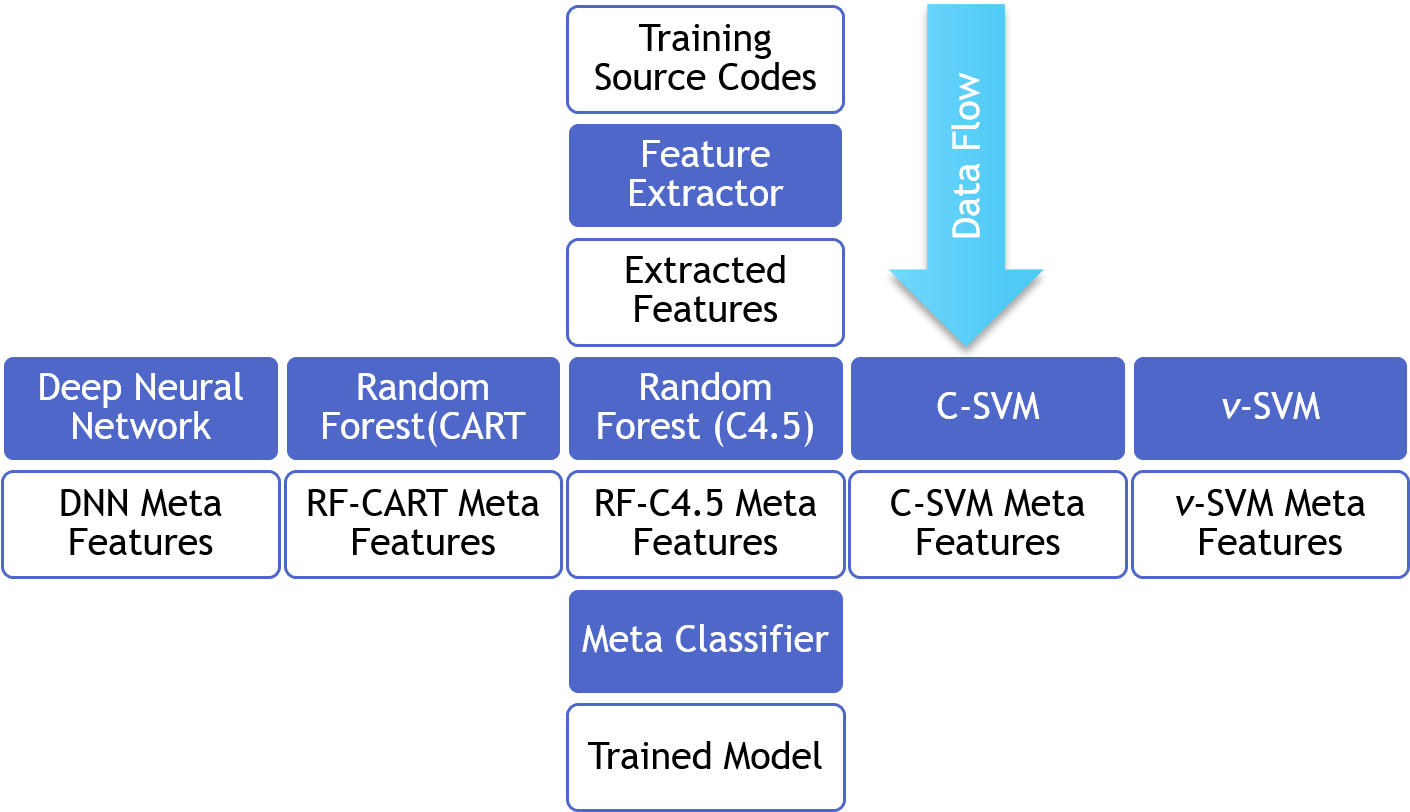}
    \caption{Block diagram of proposed author identification approach}
    \label{fig:block-diagram}
\end{figure}

Finally, to identify the author of a new source code segment, that is from the test set, the same metrics are extracted
from the test source code segments and are converted to feature vectors.
These feature vectors are fed to the meta-classifier via the base classifiers.
Using the experience from the training, the meta-classifier along with the base classifiers predict the class
labels of the test source code segments.
Figure~\ref{fig:block-diagram} shows the block diagram of the proposed approach for author identification of source
codes written by multiple authors.

In the following sub-sections, the building blocks of the author identification approach are described.

\subsection{Dataset}
\label{subsec:dataset}
Some careful considerations are needed while choosing the dataset.
Data must be collected from a diverse population of programmers and should provide enough information about the
authors so that a clear distinction can be computed from author to author and valid comparison of their programming
style can be made.
In addition, the dataset must be close to real-world data as well as open for academic study\cite{Lange}.

In our study, we have generated our dataset based on open source codes from github.com.
All the source codes have a permissive license like MIT or BSD\@.
The dataset contains 6063 python source code segments from 8 authors/ author groups which are considered as individual
classes.
Each source code segment contains roughly 226 lines on average.
Source code segments of each author are roughly split into 2:1 ratio to make the training and testing set.

Each class label consists of authors and contributors.
By author, we mean the true owner of the projects.
This could be a single author or a group of authors.
By contributors, we mean a group of people who are not the owner of the project but willingly contribute to the
project by writing or editing a segment of it.
The number of authors and the number of contributors per class-label is listed in table~\ref{table:dataset}.

\renewcommand{\arraystretch}{1.5}
\begin{table}[ht]
    \caption{Number of authors and contributors for each class}
    \label{table:dataset}
    \begin{tabular} {| p{2.9cm} | p{1.3cm} | p{1.9cm} |}
        \hline
        \textbf{Class Label} & \textbf{Number of \newline Authors} & \textbf{Number of \newline Contributors} \\
        \hline
        Azure & 3 & 136 \\
        \hline
        GoogleCloudPlatform & 33 & 820 \\
        \hline
        StackStorm & 2 & 147\\
        \hline
        dimagi & 2 & 101 \\
        \hline
        enthought & 9 & 224\\
        \hline
        fp7-ofelia & 1 & 4\\
        \hline
        freenas & 2 & 126 \\
        \hline
        sympy & 2 & 712 \\
        \hline

    \end{tabular}
\end{table}
\renewcommand{\arraystretch}{1}

\subsection{Metric Extraction}
\label{subsec:metric-extraction}
Previously, Shevertalov, Lange, Bandara, and Zhang\cite{Shevertalov,Lange,Bandara,Zhang} used source code metrics
for author identification.
From a set of probable code metrics, Lange selected the optimal set of code metrics using the genetic algorithm.
Bandara used almost the same set of source code metrics.
We have used the same set of metrics for our author identification approach only except the access modifier metric.
The access modifier feature is present only in a limited number of programming languages and makes the whole system
language dependent.
Table~\ref{table:metric} shows the set of metrics to be used and corresponding descriptions.

\renewcommand{\arraystretch}{1.5}
\begin{table}[ht]
    \caption{Set of code metrics and descriptions}
    \label{table:metric}
    \begin{tabular} {| p{2cm} | p{6cm} |}
        \hline
        \textbf{Metric Name} & \textbf{Metric Description} \\
        \hline
        Line Length &
        This metric measures the number of characters in one source code line.\\
        \hline
        Line Words &
        This metric measures the number of words in one source code line. \\
        \hline
        Comments Frequency &
        This metric calculates the relative frequency of line comment,
        block comment and optionally doc-comment used by the programmers. \\
        \hline
        Identifiers Length &
        This metric calculates the length of each identifier of programs. \\
        \hline
        Inline \newline Space-Tab &
        This metric calculates the whitespaces that occur on the interior
        areas of non-whitespace lines. \\
        \hline
        Trail \newline Space-Tab &
        This metric measures the whitespace and tab occurrence at the end
        of each non-whitespace line. \\
        \hline
        Indent \newline Space-Tab &
        This metric calculates the indentation whitespaces used at the beginning
        of each non-whitespace line.\\
        \hline
        Underscores &
        This metric measures the number of underscore characters used in
        identifiers. \\
        \hline
    \end{tabular}
\end{table}
\renewcommand{\arraystretch}{1}

After extracting the metrics, we have counted the number of occurrences for each possible values for each of the
metrics.
For example, for underscore metrics, we have counted the number of words with no underscore, one underscore, two
underscores etc.
These counts have been fed to the base classifiers.

\subsection{Base Classifiers}
\label{subsec:base-classifiers}
There are total five base classifiers in our author identification system.
They are -- DNN, random forest based on CART, random forest based on C4.5, $C$-SVM and $\nu$-SVM\@.
Each of the base classifiers is described below:

\subsubsection{Deep Neural Network}
The DNN model used as the base classifier consists of 14 layers.
Data are fed to the DNN as batches of 32 entries.
They are one input layer, followed by eight fully connected layers, a dropout layer, a fully connected layer, a
dropout layer, a fully connected layer and finally the output layer.

In the fully connected layers, \emph{ReLU} activation function and in the output layer \emph{softmax} activation
function are used.
\emph{Categorical cross-entropy} is chosen as the loss function.
\emph{Adam}\cite{Kingma} optimizer is used to optimize the network.

\subsubsection{Random Forest}
The second base classifier is a random forest with one hundred decision trees.
Classification and Regression Tree(CART)\cite{CART} algorithm is used to build the trees which selects the split
node based on Gini impurity.

The third base classifier is another random forest with one hundred decision trees.
Decision trees in the third base classifier are built with the C4.5\cite{C4.5} algorithm.
This algorithm chooses the split node based on the entropy ratio.

\subsubsection{Support Vector Machine}
The fourth base classifier is a $C$-support vector classifier.
It is a support vector machine where $C$ is a penalty parameter for the error term.

The fifth base classifier is a $\nu$-support vector classifier.
It is a support vector machine where $\nu$ is the upper bound of training error and the lower bound of the number of
support vectors.

\subsection{Meta Classifier}
\label{subsec:meta-classifiers}
We have used another deep neural network as the meta-classifier.
The outputs of the base classifiers (meta-features) are fed to the meta-classifier to learn the mapping from the
meta-features to the actual output.

The neural network consists of 19 layers.
They are one input layer, followed by eight fully connected layers, a dropout layer, two fully connected
layers, a dropout layer, a fully connected layer, a dropout layer, a fully connected layer, a dropout layer, a
fully connected layer, a dropout layer, and finally the output layer.
The output from this output layer is the final output of our author identification system for source code segment written
by multiple authors.

The activation functions of the network are \emph{ReLU} for fully connected layers and \emph{softmax} for the
output layer.
The loss function used in the meta-classifier is \emph{categorical cross-entropy}.
\emph{Stochastic Gradient Descent(SGD)} is used as the optimizer of the meta-classifier.

\begin{figure}[h]
    \centering
    \includegraphics[scale=.365, angle=0]{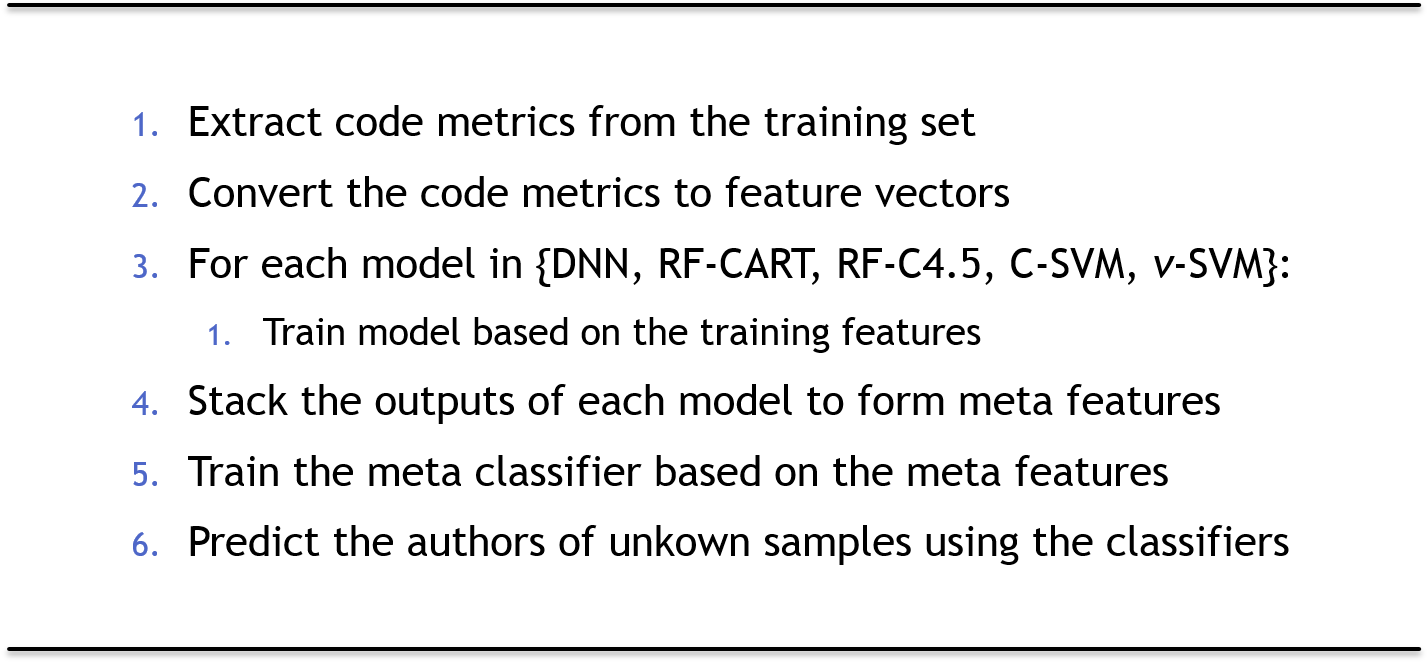}
    \caption{Steps for training the stacking ensemble system}
    \label{fig:algo}
\end{figure}

\renewcommand{\arraystretch}{1.5}
\begin{table}[ht]
    \caption{Parameter values of the classifiers}
    \label{table:parameters}
    \begin{tabular} {| p{2.5cm} | p{2.5cm} | p{2.5cm} |}
        \hline
        \textbf{Classifier} &\textbf{Parameter} & \textbf{Value} \\
        \hline
        $C$-SVM & $C$ & $1.0$ \\
        \hline
        $\nu$-SVM & $\nu$ & $0.15$ \\
        \hline
        Base DNN & learning rate & $0.01$ \\
        \hline
        \multirow{2}{*}{Adam optimizer} & $\beta_1$ & $0.9$ \\
        \cline{2-3}
        & $\beta_2$ & $0.999$ \\
        \hline
        Meta DNN & learning rate & $0.001$ \\
        \hline
        SGD optimizer & momentum & $0$ \\
        \hline
    \end{tabular}
\end{table}
\renewcommand{\arraystretch}{1}

\renewcommand{\arraystretch}{1.5}
\begin{table}[ht]
    \caption{Accuracy of the base classifiers}
    \label{table:base-results}
    \begin{tabular} {| p{5.5cm} | p{2.5cm} |}
        \hline
        \textbf{Classifier Name} & \textbf{Accuracy} \\
        \hline
        Deep Neural Network & 82\% \\
        \hline
        CART Based Random Forest & 83\% \\
        \hline
        Random Forest & 83\% \\
        \hline
        $C$-Support Vector Machine & 79\% \\
        \hline
        $\nu$-Support Vector Machine & 79\% \\
        \hline
    \end{tabular}
\end{table}
\renewcommand{\arraystretch}{1}

\renewcommand{\arraystretch}{1.5}
\begin{table*}[ht]
    \small
    \caption{Comparison among the methods for source code segment author identification}
    \label{table:result-comp}
    \begin{tabular} {| p{3.75cm} | p{2cm} | p{1cm} | p{.75cm} | p{3cm} | p{2cm} | p{2.5cm} |}
        \hline
        \rotatebox{90}{\textbf{Method Name}} & \rotatebox{90}{\textbf{Features}}
        & \rotatebox{90}{\textbf{\parbox{6em}{Language independent features}}}
        & \rotatebox{90}{\textbf{Multiple author}} & \rotatebox{90}{\textbf{\parbox{5em}{Number of classes}}}
        & \rotatebox{90}{\textbf{\parbox{6em}{Total source  code segment}}} & \rotatebox{90}{\textbf{Accuracy}} \\
        \hline
        Information retrieval \newline approach\cite{Burrows} & Character level \newline n-gram & Yes & No & 100
        & 1640 & 67\% \\
        \hline
        Code metric histogram\cite{Lange} & 7 code metrics & Yes & No & 20 & 4068 & 55\% \\
        \hline
        Genetic algorithm\cite{Shevertalov} & 4 code metrics & Yes & No & 20 & N\textbackslash A & 75\% \\
        \hline
        Deep neural network\cite{Bandara} & 9 code metrics & No & No & 10, 10, 8, 5, 9
        & 1644, 780, 475, 131, 520 & 93\%, 93\%, 93\%, 78\%, 89\%\\
        \hline
        Support vector machine\cite{Zhang} & 46 code metrics & No & No & 8, 53 & 8000, 502
        & 98\%, 80\% \\
        \hline
        \emph{Stacking ensemble method} & 8 code metrics & Yes & Yes
        & 8 (group of authors) & 6063 & 87\% \\
        \hline
    \end{tabular}
\end{table*}
\renewcommand{\arraystretch}{1}

\subsection{Training}
\label{subsec:training}
We have implemented our author identification system for source code segment written by multiple authors in multi-class
classification category.
Here, a unique list of authors(or groups of authors) of the source code segments in the training set is treated as classes.
The author identification system produces its confidence for each class of being the actual class of given source
code.
The actual author is expected to have the highest confidence.

Roughly, 67\% source code segments from each class formed the training dataset and rest are used for testing.
The training set contains 4034 files and the test set contains 2039 source code segments.

The training stage of our system is divided into three phases -- feature extraction from the source code segments,
training the base classifiers and training the meta-classifier.
Figure~\ref{fig:algo} shows the steps followed in our author identification system for source code segment written by multiple authors.

First of all, the source code metrics mentioned in table~\ref{table:metric} are extracted from source code segments.
Then the extracted metrics are converted to feature vectors as mentioned in section~\ref{subsec:metric-extraction}.
These feature vectors are fed to each of the base classifiers as input.

The base classifiers run according to their own learning algorithm to learn to identify the writing style of each
class.
During this training phase, several configurations of each of the base classifiers, specially DNN, are used to find
out which configuration works best for the training set.

After completing the training of each of the base models, the posterior probability for each input in the training
set is generated.
This produced a $5 \times |classes|$ sized feature vector for each of the input feature vectors where $|classes|$ is the
number of classes.
These feature vectors are known as meta-features.
Meta features are fed to the meta-classifier along with the class labels through which the meta-classifier learned
to predict the actual class from the meta-features.

\section{Experimental Results}
\label{sec:experimental-results}

\subsection{Experimental Setup}\label{subsec:experimental-setup}
While implementing our author identification system for source code segment written by multiple contributors, we have used
keras as the framework for deep neural networks and Scikit Learn\cite{SkLearn} as the library for
general purpose machine learning.
For data pre-processing and visualization, we have used numpy and pandas\cite{pandas} library.
We have developed a feature extractor that extracts the features mentioned in table~\ref{table:metric} from the source
codes.

For $C$-SVM, the parameter $C$ is a penalty for the error term.
For $\nu$-SVM, the parameter $\nu$ is an upper bound to the training error and lower bound to the number of support
vectors.
During the experiment, we found that for both the random forests, a hundred trees were sufficient to converge to the
highest accuracy.
After numerous iterations, we reached to a decision that the set of values stated in table~\ref{table:parameters}
classifies the source code segments most accurately.

\emph{Accuracy} and \emph{f1-score} were used to evaluate the accuracy of our method.
\emph{Accuracy} is the ratio between the number of correctly identified samples and the number of total samples.
\emph{F1-score} is the harmonic mean of \emph{precision} and \emph{recall}.
\emph{Micro averaging} was used to compute the \emph{f1-score}.

\subsection{Results of The Base Classifiers}\label{subsec:results-base-classifiers}
Table~\ref{table:base-results} contains the accuracies for the five base models of our stacking ensemble method.

\subsection{Results of The Meta Classifier}\label{subsec:results-meta-classifier}
After training the meta-classifier by the meta-features, we have achieved 87\% accuracy with f1-score $0.86$.
Identifying the authorship of source codes is more difficult when the number of authors is more than one
as the writing style of the source code is then inconsistent from segment to segment.
Table~\ref{table:result-comp} shows a comparison between the type of features, language independence, the capability
of handling multiple authorship, number of classes and the total number of source code segments used in training and
testing.
From that table, we can see that even after dealing with source code segments written by multiple authors, our method has
achieved an accuracy that is pretty close to that of the methods that deal with single authors.
Our chosen set of metrics is compact and is still able to achieve a satisfactory accuracy.
Alongside a number of works suffer from choosing a set of metrics that are not language independent.
So, the main contribution of this work is the identification of multiple authors using language independent set of
metrics.

\section{Conclusion}
\label{sec:conclusion}
Here, we have proposed a new approach for identifying the author of source code segment where the number of authors of
the source code segment is more than one.
The main challenge of this work is to select the base estimators from a large number of possible combinations.
Again, as several classifiers need to be trained, each classifier needs to be fine-tuned individually to
produce a good final result.
On the other hand, the problem of identifying the authorship of source code segments is harder when the number of authors is
more than one.

We have developed a stacking ensemble classifier that consists of five base classifiers and a meta-classifier which
uses a relatively small set of code metrics that are relatively easy to compute and language independent as
well.

In spite of the fact that our stacking ensemble method achieved a satisfactory accuracy, this still can be improved.
Even though our code metrics are language independent, we only tested with python source code segments.
Future works may test on other languages and check how the set of metrics works for other languages.
Other sets of metrics can also be examined to see how they contribute to the writing style of source code segments.


\vspace{12pt}
\end{document}